\documentclass[useAMS,usenatbib]{mn2e}

\usepackage {graphicx}
\usepackage {times}
\def\aj{\,{AJ}}
\def\apj{\,{\rm ApJ}}
\def\apjl{\,{\rm ApJL}}

\def\mnras{\,{\rm MNRAS}}
\def\kms{\, km s$^{-1}$ }

\title[The Origin of E+A Galaxies]{266 E+A Galaxies Selected from the Sloan Digital Sky Survey Data Release 2: The Origin of E+A Galaxies}
\author[T. Goto]
{Tomotsugu Goto$^{1}$\thanks{E-mail:tomo@jhu.edu}
  \\
  $^{1}$ Department of Physics and Astronomy, The Johns Hopkins
  University, 3400 North Charles Street, Baltimore, MD 21218-2686, USA
}

\begin{document}
\maketitle


%




\begin{abstract}\label{abstract}

 E+A galaxies are characterized as a galaxy with strong Balmer
 absorption lines but without any [OII] nor H$\alpha$ emission
 lines. The existence of strong Balmer absorption lines indicates that E+A
 galaxies have experienced starburst within recent one Gyr. However, the
 lack of [OII] and H$\alpha$ emission lines indicates that E+A galaxies
 do not have any on-going star formation. Therefore, E+A galaxies are
 interpreted as a post-starburst galaxy. For many years, however, it has been a mystery 
 why E+A galaxies started starburst and why they quenched the star formation
 abruptly. 
 Using  one of the largest samples of 266 E+A galaxies carefully selected from the Sloan Digital Sky Survey Data Release 2, we have investigated the environment of E+A galaxies from 50 kpc to 8 Mpc scale, i.e., from a typical distance to satellite galaxies to the scale of large scale structures. We found that E+A galaxies have excess of local galaxy density only at a scale of $<100$ kpc (with a two $\sigma$ significance), but not at the cluster scale ($\sim$1.5 Mpc) nor in the scale of large scale structure ($\sim$8 Mpc). These results indicate that E+A galaxies 
are not created by the physical mechanisms associated with galaxy clusters or the large scale structure, but are likely to be created by the dynamical interaction with closely accompanying galaxies at a $<$100 kpc scale. The claim is also supported by the morphology of E+A galaxies. We have found that almost all E+A galaxies have a bright compact core, and that $\sim$30\% of E+A galaxies have dynamically disturbed signatures or the tidal tails, being quite suggestive of morphological appearance of merger/interaction remnants.

\end{abstract}

\section{Introduction}\label{intro}

  Dressler \& Gunn (1983; 1992) found galaxies with mysterious spectra while
 investigating high redshift cluster galaxies.
 The galaxies had strong Balmer absorption lines with no
 emission in [OII]. These galaxies are called 
 ``E+A''
 galaxies since their spectra looked like a superposition of that of
 elliptical galaxies (Mg$_{5175}$, Fe$_{5270}$ and Ca$_{3934,3468}$
 absorption lines) and that of A-type stars (strong
 Balmer absorption lines).
  The existence of strong Balmer absorption lines shows 
 that these galaxies have experienced starburst recently (within a
 Gyr; Goto 2004b).  However, these galaxies do not show any sign of on-going star
 formation as non-detection in the [OII] emission line
 indicates.  
   Therefore, E+A galaxies are interpreted as a post-starburst galaxy,
 that is,  a galaxy which truncated starburst suddenly (Dressler \& Gunn 1983,
 1992; Couch \& Sharples 1987; MacLaren, Ellis, \& Couch 1988; Newberry
 Boroson \& Kirshner 1990; 
 Fabricant, McClintock, \& Bautz 1991; Abraham et al. 1996).
  The reason why they started starburst, and why they abruptly stopped starburst remains one
 of the mysteries in galaxy evolution. 

   One possible explanation for the E+A phenomena is a dust enshrouded starburst galaxy. In this scenario, E+A galaxies are not post-starburst galaxies, but in reality star-forming galaxies whose emission lines are invisible in optical wavelengths due to the heavy obscuration by dust (Smail et al. 1999; Poggianti \& Wu 2000).  
   A straightforward test for these scenarios is to observe in radio wavelengths
  where the dust obscuration is negligible.
    Miller \& Owen (2001) observed radio continua of 15 E+A galaxies and
  detected moderate levels of star formation in only 2 of them.
 Goto (2004b) performed 20 cm radio continuum observation of 36 E+A galaxies using the VLA, and found that none of their E+A galaxies are detected in 20 cm, suggesting that E+A galaxies are not dusty-starburst galaxies.

Alternative explanations to E+As include cluster related physical mechanisms.  Since E+A galaxies were found in cluster regions at first, both in low  redshift clusters (Caldwell et al.    1993, 1996;  Caldwell \& Rose 1997; Castander et al. 2001; Rose et
  al. 2001; Poggianti et al. 2004) and high redshift   clusters (Sharples et   al. 1985; Lavery \& Henry 1986; Couch \&  Sharples 1987; Fabricant,  McClintock, \& Bautz 1991; Belloni   et al. 1995; Barger et al. 1996; Fisher et  al. 1998; Morris et al. 1998; Couch et al. 1998;   Dressler et al. 1999; Tran et al. 2003,2004), a cluster specific phenomenon was thought to be responsible for the
  violent star formation history of E+A galaxies. For example, a ram-pressure
  stripping model (Spitzer \& Baade 1951, Gunn \& Gott 1972, Farouki \&
  Shapiro  1980; Kent 1981; Abadi, Moore \& Bower 1999; Fujita \& Nagashima 1999;
 Quilis, Moore \& Bower 2000; Fujita 2004; Fujita \& Goto 2004)
  may first accelerate star formation of cluster galaxies and later turn it
  off as well as tides from the cluster potential (e.g., Fujita 1998) and the evaporation of the cold gas (e.g., Fujita 2004).

 Another possible origin is the galaxy-galaxy interaction, which has been known to trigger
      star formation in a pair of galaxies (Schweizer 1982; Lavery \&
      Henry 1988; Liu \& Kennicutt 1995a,b; Schweizer 1996).
 Bekki, Shioya, \& Couch (2001) modeled galaxy-galaxy mergers with
    dust extinction, confirming that such systems can produce
    spectra which evolve into E+A spectra.

 Although these scenarios are all plausible, the definitive
 conclusion has not been drawn yet, mainly because E+A galaxies are a vary rare class of galaxies and have been difficult to study in a large number. 
 The largest sample of E+A galaxies to date is  presented by Galaz (2000), which, however, is a heterogeneous sample  of only 50 E+As.
 Below, we overcome this problem by selecting E+A galaxies from the $\sim$250,000 spectra of the Sloan Digital Sky Survey Data Release 2 (SDSS; Abazajian et al. 2004).
 Unless otherwise stated, we adopt the best-fit WMAP cosmology:
 $(h,\Omega_m,\Omega_L) = (0.71,0.27,0.73)$
 (Bennett et al. 2003).

\begin{figure}
\begin{center}
\includegraphics[scale=0.5]{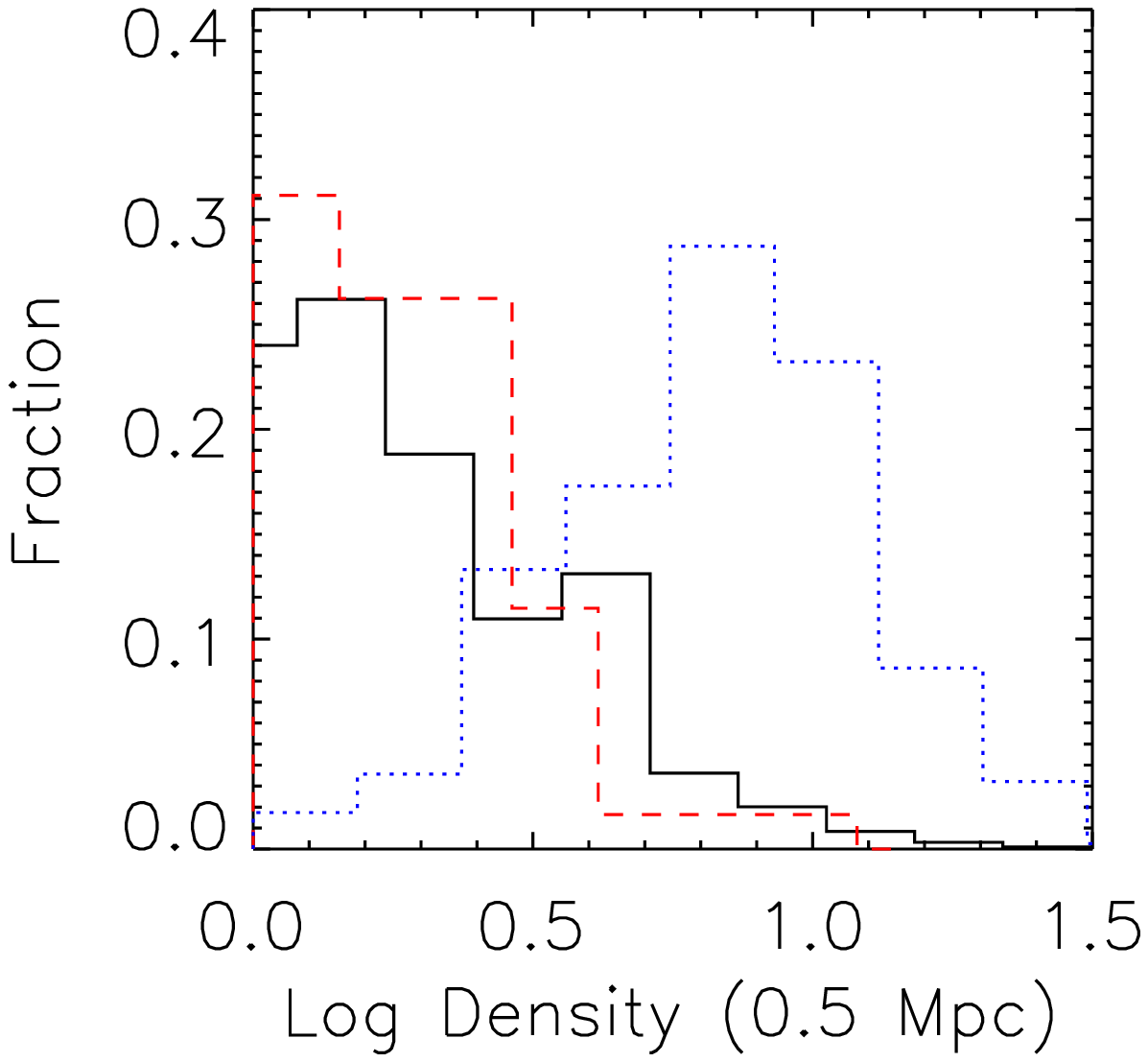}
\includegraphics[scale=0.5]{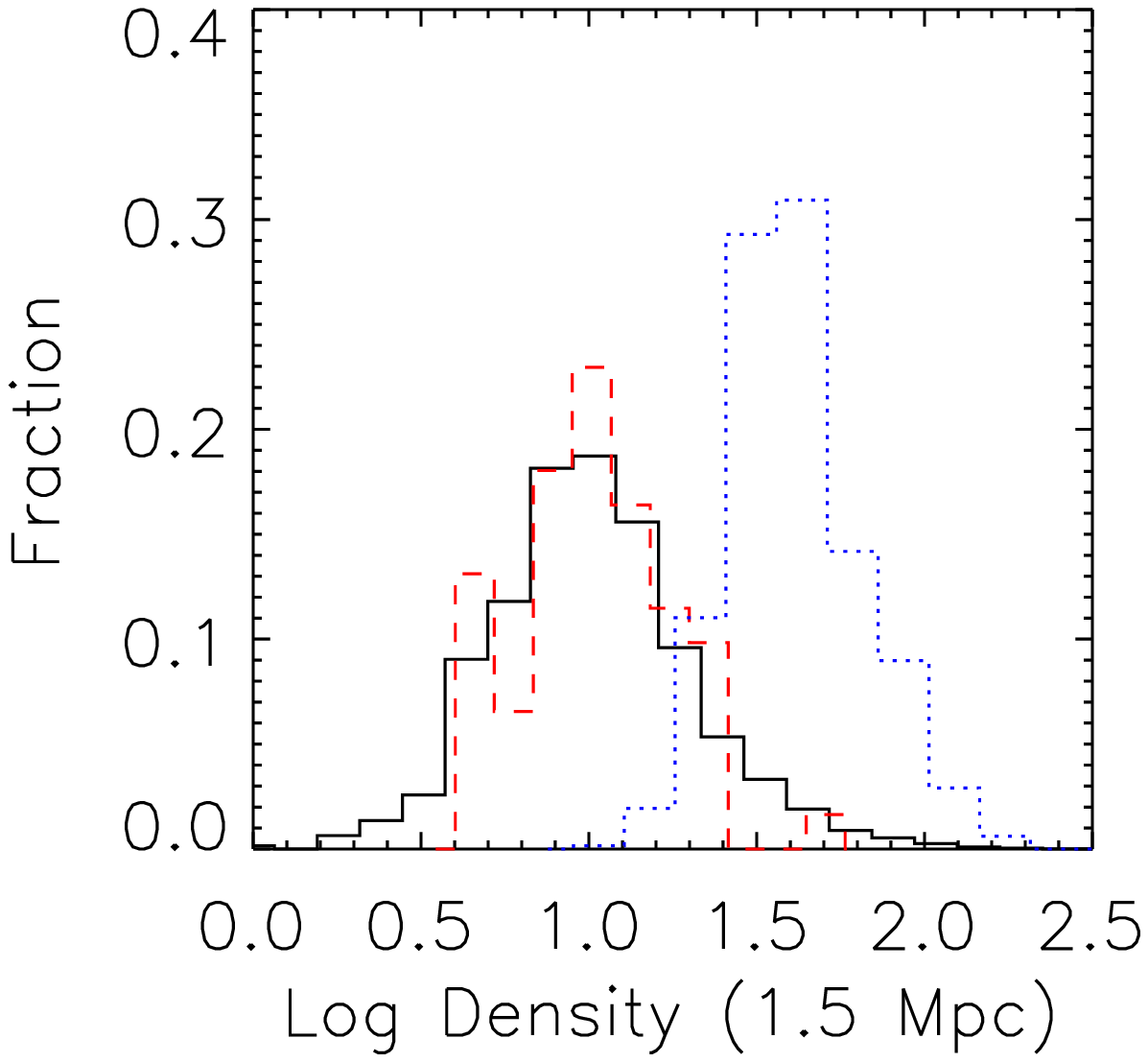}
\includegraphics[scale=0.5]{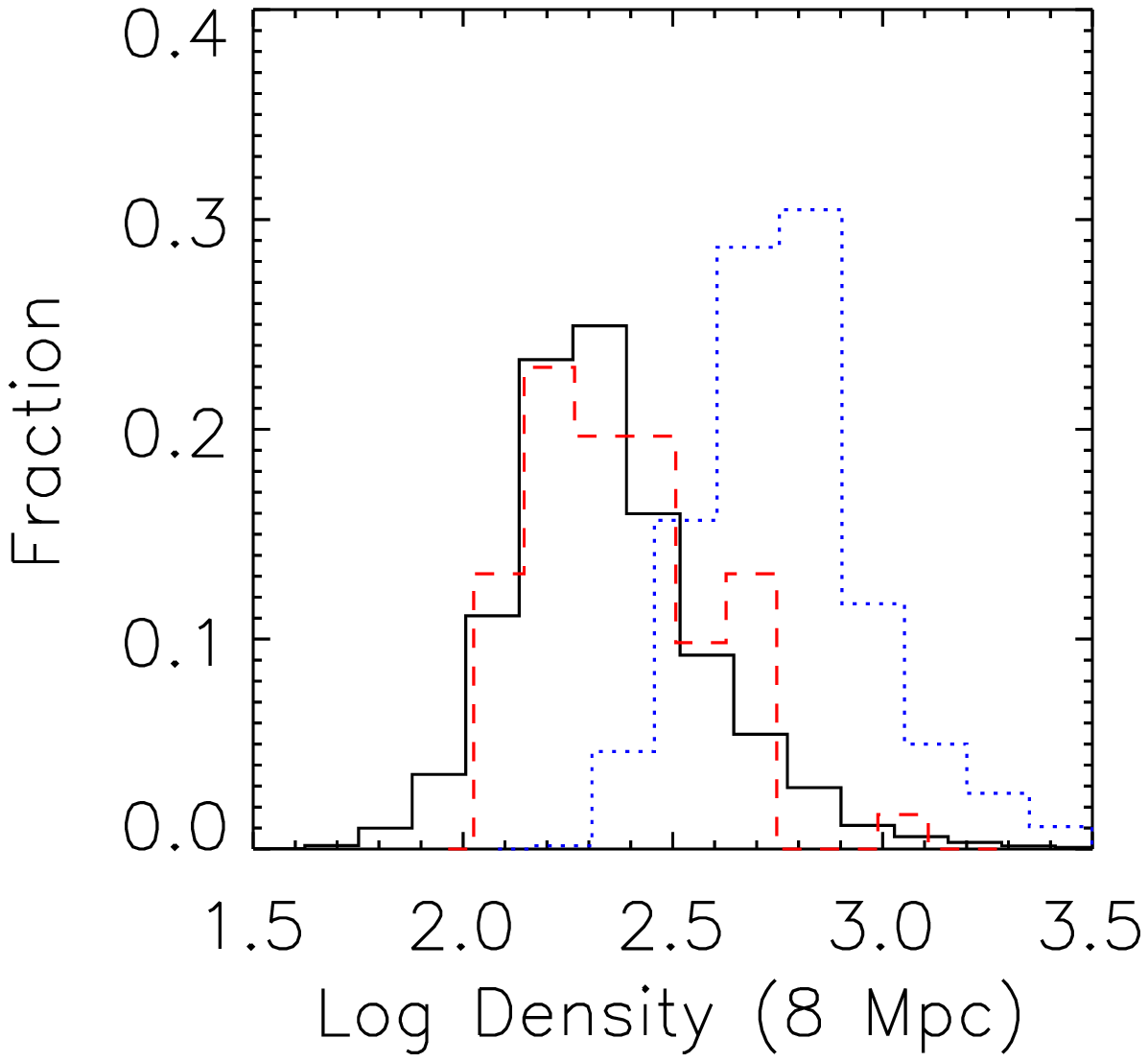}
\end{center}
\caption{Normalized Distributions of Local Galaxy Density. The local galaxy densities are measured within 0.5 (top panel), 1.5 (middle panel) and 8 (bottom panel) Mpc of radius in angular direction and $\pm$1000 km s$^{-1}$ in the line of sight direction within the volume limited sample ($z<0.16$ and $Mz<-21.84$). The solid, dashed, dotted lines denote field, E+A, and cluster galaxies, respectively.  
}\label{fig:0.5-8Mpc}
\end{figure}

\section{Data \& E+A Selection}\label{data}
  
 We have created a new sample of E+A galaxies using the publicly available SDSS DR2 (Abazajian et al. 2004).
 We only use those objects classified as galaxies ({\tt type=3}, i.e., extended) with spectroscopically measured redshift of $z>0.01$. These two criteria can almost entirely eliminate contamination from mis-classified stars and HII regions in nearby galaxies. We further restrict our sample to those with spectroscopic signal-to-noise $>$10 per pixel (in continuum of the $r-$band wavelength range) since it is difficult to measure absorption lines when signal-to-noise is smaller than 10. 
 For these galaxies, we have measured H$\delta$,
  [OII] and H$\alpha$ equivalent widths (EWs) and obtained their errors
  using the flux summing  method similar to the one described in Goto et al. (2003a). 
 The only difference from Goto et al. (2003a) is that we only used the wider window of  4082$-$4122\AA\ to measure the H$\delta$ line to ease the comparison with models (c.f. Goto et al. 2003a combined the narrower window and took the maximum). 
 Once the lines are measured, we have selected E+A galaxies as those with H$\delta$ EW $>$ 5.0\AA, and H$\alpha$ EW $>$ -3.0\AA, and [OII] EW $>$ -2.5\AA\footnote{Absorption lines have a positive sign throughout this paper.}. Although the criteria on emission lines allow small amount of emission in the E+A sample, they are relatively small amount in terms of the SFR.   We also exclude galaxies at $0.35<z<0.37$ from our sample since the overlapping sky feature at 5577\AA\ disturbs the H$\delta$ EW measurement.
 Those criteria are empirically determined to include only secure E+As, excluding suspicious ones. However in general, our criteria are more strict than previous ones (e.g., H$\delta$ EW $>$ 4.0\AA\ and [OII] EW $>$ -2.5\AA).
 We stress the advantage in having the information on the H$\alpha$ line in selecting E+A galaxies. Previous samples of E+A galaxies
 were often selected based solely on [OII] emission and Balmer
 absorption lines either due to the high redshift of the samples or due to
 instrumental reasons. According to Goto et al. (2003a), such selections of E+A
 galaxies without information on H$\alpha$ line would suffer from 52\%
 of contamination from H$\alpha$ emitting galaxies, whose morphology and color are very different from that of E+A galaxies (Goto et al. 2003b; Goto 2003; G03 hereafter).
 Blake et al. (2005) selected E+A galaxies from the 2dF using only Balmer and [OII] lines to find that some E+A galaxies in their sample have the H$\alpha$ line in emission. 
  Among $\sim$250,000 galaxies which satisfy the redshift and S/N
  cut with measurable [OII], H$\delta$ and H$\alpha$ lines, we have found 266 E+A galaxies, which is one of the largest sample of E+A galaxies. 
  The 266 E+A galaxies span a redshift range of  0.032 $\leq z \leq$ 0.342, within which H$\alpha$ line is securely covered by the SDSS spectrograph. Although this range covers about 2 Gyr of lookback time, majority of our E+A galaxies lie at $z\sim 0.1$. Therefore, we do not consider evolution within the E+A sample in this paper.

\section{Results}\label{sec:Results}

\subsection{Mpc Scale Environment}\label{sec:Mpc}

 In order to test if E+A galaxies are created by the cluster related mechanisms or not, we first investigate environment of E+A galaxies in Mpc scale. If the origin of E+A galaxies is  physical mechanisms related to galaxy clusters or large scale structures, E+A galaxies should show excess in abundance in such environments. We quantify the environment of E+A galaxies by measuring the local galaxy density around E+A galaxies: First, we limit our sample galaxies to the volume limited sample of $0.01<z<0.16$ and $M_r<-21.84$. These criteria are determined to include maximum number of 61 E+A galaxies (out of 266) in the sample. Next, we count the number of neighbors within $\pm 1000$ \kms in the line of sight direction, and 0.5, 1.5, and 8.0 Mpc radius in angular direction. The number of neighbors are divided by the angular area they subtend to be converted to the (surface) local galaxy density in 0.5, 1.5, and 8.0 Mpc scales. 
 Figure \ref{fig:0.5-8Mpc} shows the local galaxy density distribution of E+A galaxies (dashed line), and all galaxies (solid line) in the volume limited sample.
 The local galaxy density is measured in 0.5, 1.5, and 8.0 Mpc scales in the top, middle, and bottom panels, respectively. 
As a reference, we plot the density distributions of cluster galaxies found in Goto et al. (2004c; see also Goto et al. 2002a,b for a cluster catalog) in the dotted line. In Fig. \ref{fig:0.5-8Mpc}, it is recognized that cluster galaxies have very different density distribution from the E+A and the field galaxies, with its peak at much denser regions. A Kolomogorov-Smirnov test shows that the distributions of cluster galaxies and E+A or field galaxies are different at a $>$99.9\% significance level at all scales. 
 On the other hand, the density distribution of E+A galaxies and the field galaxies are indistinguishable at all scales.
 These results indicate that E+A galaxies are not likely to be created by physical processes at work at cluster scale (0.5, 1.5 Mpc), nor by the large scale structure (8 Mpc). We discuss the relevance to the previous work and more physical implications in Section \ref{discussion}.

\subsection{kpc Scale Environment}\label{sec:kpc}

\begin{figure}
\begin{center}
\includegraphics[scale=0.4]{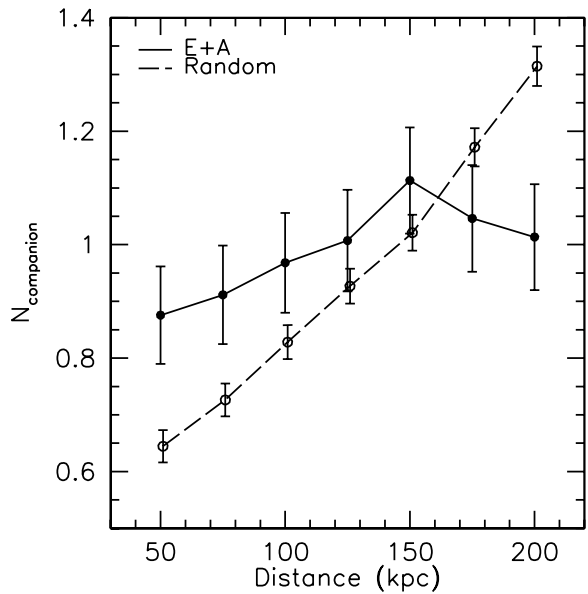}
\end{center}
\caption{Number of companion galaxies within a 50-200 kpc radius. The solid line is for E+A galaxies and the dashed line is for random (field) galaxies. The number of companion galaxies are counted within $-23.0<M_r<-20.25$ after k-correction. The fore/background counts are statistically subtracted. 
}\label{fig:kpc_raw}
\end{figure}

\begin{figure}
\begin{center}
\includegraphics[scale=0.4]{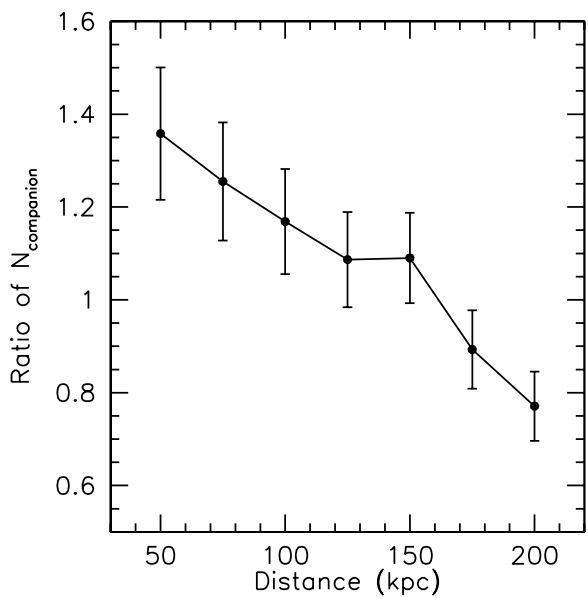}
\end{center}
\caption{The number ratio of companion galaxies of E+As to that of random (field) galaxies. 
}\label{fig:kpc_ratio}
\end{figure}

Next, we investigate the environment of E+A galaxies in much smaller physical scale of 50-200 kpc. In the previous subsection, we used the volume limited sample of spectroscopic galaxies to investigate Mpc scale environment. However, this method cannot be used to probe kpc scale environment since (i) the SDSS fiber spectrograph cannot be located closer than 55 arcsec. And thus, significant fraction of galaxies within 55 arcsec from the central E+A galaxies are missing from the SDSS spectroscopic sample; (ii) the spectroscopic sample has a bright flux limit ($r=17.77$), and thus, cannot probe the over-density of faint galaxies at a small scale. We overcome these problems by utilizing the SDSS imaging data, which provide relatively good star/galaxy separation as faint as $r=21.5$ mag (Scranton et al. 2002). At several radii between 50 and 200 kpc, we count number of galaxies in the imaging data between $M_r=-23$ and $-20.25$ after applying k-correction. This  k-correction was applied on
galaxy-by-galaxy basis assuming the redshift of the central E+A galaxy using the code provided by Blanton et al. (2003; v3\_2). This magnitude range allows us to use all 266 E+A galaxies including the furtherest one at $z=0.342$.  The number count of background galaxies computed from randomly selected 10 deg$^2$ are adjusted to the sky area of the radii and subtracted using the same magnitude range. This background subtraction is relatively small (only $\sim$3\% in 50kpc and 30\% in 200 kpc) due to the small angular area probed, justifying our use of the imaging data to investigate kpc scales.  And, since the background is computed using a much larger part of the sky (thus more stable), we regarded errors associated with this background subtraction negligible compared with the Poisson errors associated with the sample itself.
 For a comparison, we perform the same analysis to randomly selected $\sim 2000$ galaxies in a similar redshift and signal-to-noise ratio range.
Figure \ref{fig:kpc_raw} shows the resulting average number of companion galaxies within 50-200 kpc radii around E+As (solid line) and the comparison sample (dashed line; we call this as field galaxies hereafter). To clarify the difference, we plot the number ratio of E+A companions and field companions in Fig. \ref{fig:kpc_ratio}. In both figures, error bars are based on Poisson statistics, and mainly dominated by the small number statistics of E+A galaxies. It can be seen that E+A galaxies has excess number of companion galaxies at a small scale of 50-100 kpc, with the difference increasing with decreasing radius. Although error bars on E+A sample is not so small as the field sample, the difference at 50kpc is at $\sim 2\sigma$ significance with E+As having 35\% more companion galaxies than field galaxies.
 This may suggest that the origin of E+A galaxies is associated with the excess number of companion galaxies, possibly linking to frequent merger/interaction with companions.  We further discuss this in Section \ref{discussion}. It is also worth noting that at 175-200kpc, E+A galaxies have smaller number of companion galaxies than the field galaxies. If true, this may reflect depletion of companion galaxies at this scale due to the central E+A's cannibalism of companion galaxies. In the top panel of Fig.\ref{fig:0.5-8Mpc}, we see a possible skew of the E+A's local galaxy density distribution toward lower density in 0.5 Mpc scale. This may be consistent with the results with 175-200kpc, suggesting E+A galaxies may have under-density at 175-500kpc scale. Both of the results are consistent with the hypothesis that E+A galaxies do not exist in cluster environment.

 Note that the cosmic variance is not worrisome here since both the E+A and the field samples are drawn from the large sky area covered by the SDSS ($>$3000 deg$^2$). Also note that since we have used the same background subtraction for both the E+A and the field samples, the difference between these two must stem from the difference between these two samples, not from the  background subtraction.

\begin{figure*}
\begin{center}
\includegraphics[scale=0.8]{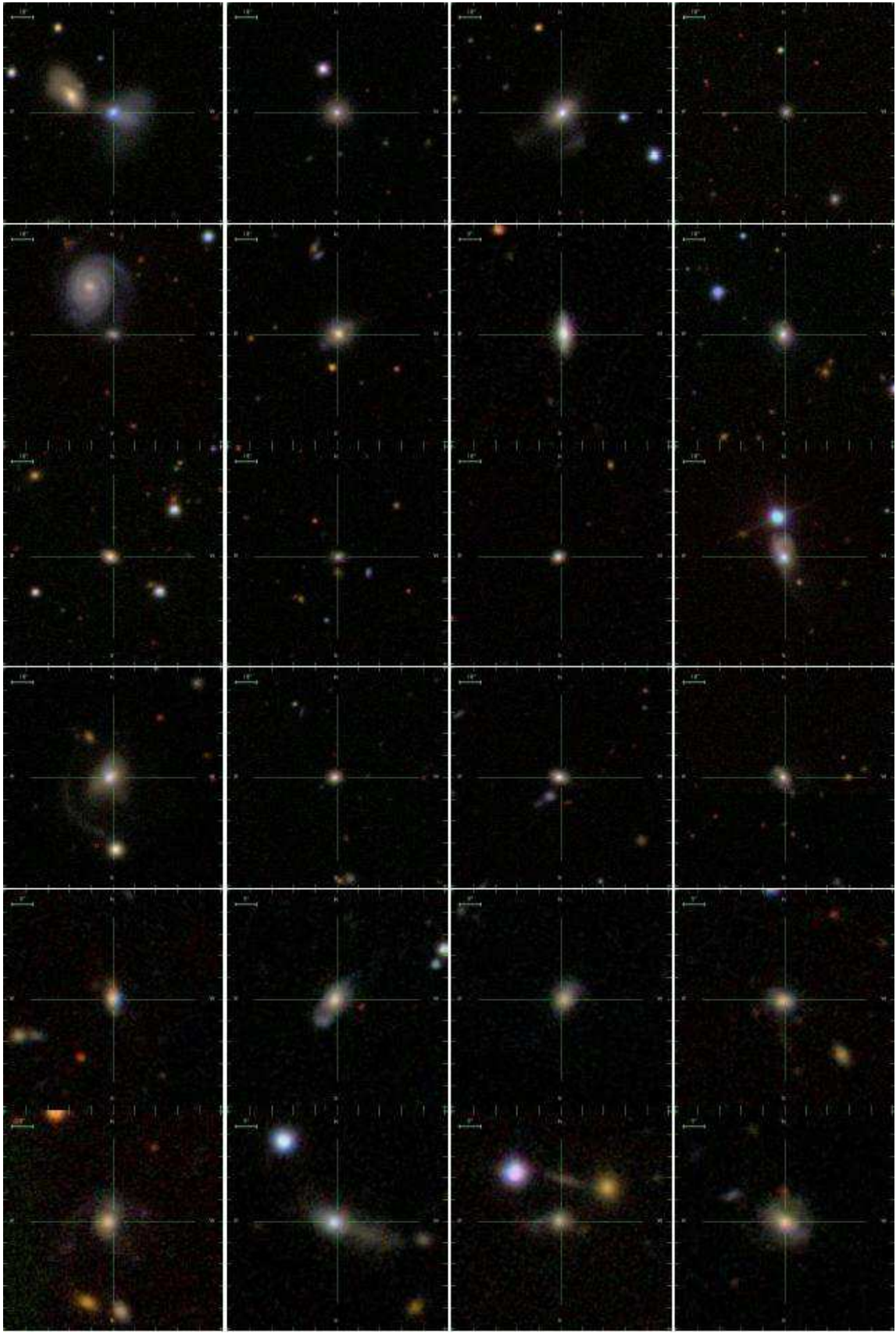}
\end{center}
\caption{Examples of $g,r,i$-composite images of E+A galaxies with H$\delta$ EW $>$7\AA. The images are sorted from low to high redshift. Only 24 lowest redshift E+As are shown. The corresponding spectra with name and redshift are presented in Fig. \ref{fig:ea7_spectra}. (The spectrum of the same galaxy can be found in the same column/row panel of Fig.\ref{fig:ea7_spectra}.)
}\label{fig:ea7_images}
\end{figure*}

\begin{figure*}
\begin{center}
\includegraphics[scale=0.9]{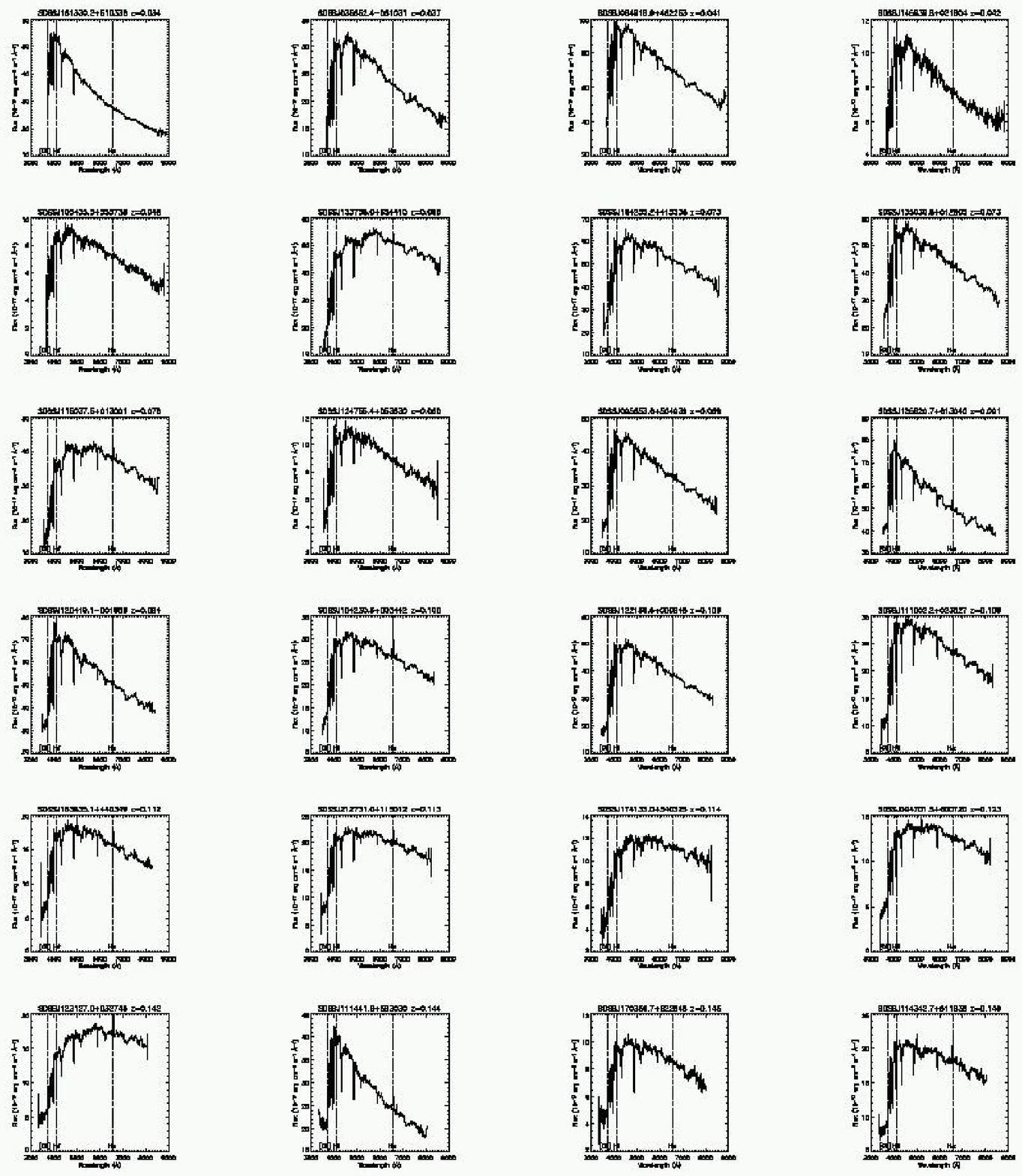}
\end{center}
\caption{Example spectra of 24 lowest redshift E+As with H$\delta$ EW $>$7\AA. The spectra are sorted from low redshift. Each spectrum is shifted to the restframe wavelength and smoothed using a 20 \AA\ box. The corresponding images are shown in Fig. \ref{fig:ea7_images}. (The image of the same galaxy can be found in the same column/row panel of Fig.\ref{fig:ea7_images})
}\label{fig:ea7_spectra}
\end{figure*}

\subsection{Implication from     Morphology of E+A Galaxies}

 In the previous sections, we have investigated kpc to Mpc environment of E+A galaxies, and  found that E+A galaxies have excess number of companion galaxies at $<$100 kpc scale, but not in any larger scale (Figs. \ref{fig:0.5-8Mpc} and \ref{fig:kpc_ratio}). The scale of $\sim$ 100 kpc is comparable to the distance between the Milky Way Galaxy and its satellite, Small Magellanic Cloud ($\sim$60 kpc) although the Small Magellanic Cloud is much fainter than the magnitude range we probed ($-23.0<M_r<-20.25$).  Due to the small distance, this scale may well be probed by examining the atlas  image of central E+A galaxies. In Fig. \ref{fig:ea7_images}, we show example $g,r,i$-composite atlas images of 24 lowest redshift E+A galaxies with H$\delta$ EW $>$ 7\AA. 
According to Goto (2004a), E+A galaxies with larger H$\delta$ EW are either close to the truncation epoch or with a stronger starburst.
The images are sorted from low redshift to high redshift. The corresponding spectra with names are shown in Fig. \ref{fig:ea7_spectra}. As can be seen in Fig. \ref{fig:ea7_images}, almost all E+A galaxies have very bright central core with high concentration, often in blue color. At the same time, E+As often lack a clear sign of disk. This can be also recognized in Fig. \ref{fig:concent} where we plot the restframe $u-r$ color against the concentration parameter, $Cin$, which is the  ratio of Petrosian 50\% light radius to Petrosian 90\% light radius in $r$ band  (Shimasaku et al. 2001; Strateva et al. 2001). The contours show the distribution of all (field) galaxies and the diamonds denote E+A galaxies. It can be seen that most of E+As have $Cin<0.4$, indicating that E+A galaxies have as high concentration ratio as ellipticals.  In addition, in Fig. \ref{fig:ea7_images}, there are notable number of E+A galaxies with disturbed/tidal tail signatures: For example, the lowest redshift one, SDSSJ161330.24+510335.6, has an apparent head-to-head interaction with a companion; SDSSJ084918.96+462252.6, SDSSJ125820.64+613039.6, SDSSJ120418.96-001855.8, and SDSSJ111441.52+593029.8 have low surface brightness but obvious tidal features; SDSSJ122156.4+000948.2 has an apparent companion with tidal tails; SDSSJ105435.51+555738.1 is apparently a satellite galaxy of a grand nearby spiral. Beside this one, we also found a few other cases where E+As are a satellite of a giant galaxy. These findings of apparent companion/tidally disturbed features are consistent with our findings of the excess number of companion galaxies at $<$100 kpc scale, together indicating that dynamical interaction with close companion galaxies may be the origin of E+A galaxies.

\section{Discussion}\label{discussion}

\subsection{Are E+As Cluster Related Phenomena?}
 
 In section \ref{sec:Results}, we showed that the 
 distribution of E+A galaxies is significantly different from that of cluster galaxies in 0.5-8 Mpc scales, and that E+A galaxies are ubiquitous in the field region. These results suggest that E+A galaxies are not created by the cluster related physical phenomena nor by the large scale structure. In the literature, there are some recent work which found consistent results with ours; Quintero et al. (2004) reported that K+A galaxies do not lie in high-density
 regions using a K/A star ratio to select $\sim$1000 K+A galaxies from the SDSS data (However, note that they did not apply constraints on emission lines, and thus, their K+A sample includes significant amount of H$\alpha$ emitting galaxies whose properties are different from true E+As, see G03); Mateus \& Sodr{\'e} (2004) also found that short-starburst galaxies are ubiquitous in the field regions; GO3 also showed a preliminary evidence of field E+A galaxies using the 5th nearest neighbor analysis with a smaller number of E+As. 
  During the refereeing process of this paper, Blake et al. (2005) reported that E+A galaxies found in the 2dF survey predominantly lie in the field regions, again in agreement with our result in Fig \ref{fig:0.5-8Mpc}. 

 Historically, since E+A galaxies were found in cluster regions at first, a
  cluster specific phenomenon was thought to be responsible for the
  creation of E+A galaxies such as the ram-pressure stripping or the tidal interaction with the cluster potential. 
  However, considering these recent results which found numerous E+A galaxies in the field regions, at the very least, it is clear that these field E+A galaxies cannot be explained by a physical mechanism that works only in the cluster region (e.g., the ram-pressure stripping). E+A galaxies have been often thought to be  transition objects during the cluster galaxy evolution such as the  Butcher-Oemler effect (e.g.,  Butcher \& Oemler 1984; Goto et al. 2003c  and see the references therein), the morphology-density  relation (e.g., Postman \& Geller 1984; Dressler et al. 1997; Goto et al. 2003d; Goto et al. 2004a and see the references therein), and the correlation between star formation rate and the environment (e.g., Hashimoto et al. 1998; Tanaka et al. 2004 and see the references therein). However, explaining cluster galaxy evolution using E+A galaxies may not be realistic anymore. Passive spiral galaxies (Couch et al. 1998;  Poggianti et al. 1999; Goto et al. 2003e; Yamauchi \& Goto 2004) may be an alternative candidate for the transition objects in cluster galaxy evolution instead of E+As.

\subsection{Merger/Interaction Origin of  E+A Galaxies} 

\begin{figure}
\begin{center}
\includegraphics[scale=0.6]{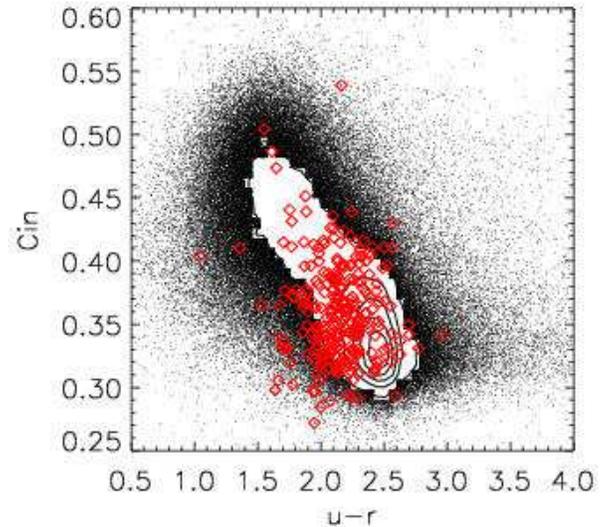}
\end{center}
\caption{The inverse of a concentration parameter, $Cin$, is plotted against restframe $u-r$ color. The contour and small dots represent distribution of all (field) galaxies. The diamonds are E+A galaxies. 
}\label{fig:concent}
\end{figure}

 In Section \ref{sec:Results}, we also  found that E+A galaxies have excess number of companion galaxies at a scale of $<100$ kpc. The atlas images of E+A galaxies show frequent signatures of companion/tidal tails (Fig. \ref{fig:ea7_images}). The findings suggest that E+A galaxies may be created by the dynamical interaction with close ($\sim 100$ kpc scale) companion galaxies.
  Previous work on this subject was based on only a few E+A galaxies due to their rarity, nevertheless, some supportive evidence can be found in the literature:  Oegerle, Hill, \& Hoessel(1991)  observationally found a tidal feature in a nearby E+A galaxy G515 (see also Carter et
 al. 1988). 
  Schweizer (1996) found several nearby E+As that have highly disturbed
 morphologies consistent with the products of galaxy-galaxy mergers.
  One E+A galaxy observed with the VLA has clear tidal tails (Chang
 et al. 2001), which support the galaxy-galaxy interaction
 picture for E+A formation.
   Poggianti \& Wu (2000) reported that  proportion of close mergers are very
 high among their e(a) sample.  Liu \& Kennicutt (1995a,b) also found E+A galaxies among 40
 merging/interacting systems they observed.
   Bartholomew et al. (2001) and Norton et al. (2001) found that
 star formation in E+A galaxies is centrally concentrated, arguing that
 their observational results are consistent with the tidal interaction
 origin, considering the fact that Moss \& Whittle (1993,2000) reported
 that early-type tidally distorted spiral galaxies are often found with
 compact nuclear H$\alpha$ emission. This discussion is also consistent with our findings of high concentration of E+A galaxies (Fig. \ref{fig:concent}). 
   Theoretically,  Bekki et al. (2001) showed that galaxy-galaxy mergers with
    high infrared luminosity  can produce e(a) spectra which evolve
 into E+A spectra (See also Shioya et al. 2001,2002).  
 Recently, based on the Supercosmos Sky Survey plates, Blake et al. (2005) found that a small but significant number of E+A galaxies have evidence of recent major galaxy mergers, such as disturbed morphologies, coalescing disks and tidal tails or envelopes.

 These previous results were mostly based on only a few nearby E+A galaxies (except Blake et al. 2005), and thus lacked statistical significance. However, they are all consistent with our findings of excess number of companion galaxies at $<$100 kpc scale based on 266 E+A galaxies, supporting the merger/interaction origin of E+A galaxies.

 Finally it is important to ponder on the difference between dynamical timescale and spectral timescale. It has been known the timescale of spectral E+A phase is $\sim$1 Gyr due to the life time of A-type stars. The dynamical timescale of galaxy-galaxy merging is more uncertain, but it is estimated to be $\sim$0.5 Gyr (Mihos 1995), and thus, perhaps shorter than the spectral timescale. This difference in timescale may be the reason why we found morphological signatures in only 7 E+A galaxies out of 24.
   This fraction is only $\sim$30\% (Fig.\ref{fig:ea7_images}), which, however, is higher than the percentage reported in the literature (e.g., $\sim$13\% in Blake et al. 2005). According to the population synthesis modeling, E+A galaxies with larger H$\delta$ EWs are closer to the truncation epoch (see Fig.2 of Goto et al. 2004b). Our selection of E+A galaxies with large H$\delta$ EWs may have resulted in the frequent merging signatures in Fig.\ref{fig:ea7_images}.

\section{Conclusions}\label{conclusion}

 E+A galaxies have been interpreted as post-starburst galaxies based on their spectral features  (strong Balmer absorption with no [OII] nor H$\alpha$ emission lines). However, the origin of the vigorous change in their star formation history has been a mystery mainly due to the lack of a statistical sample. Using the large sample of 266 E+A galaxies carefully selected from the SDSS DR2, we have found the following:

\begin{itemize}
 \item Local galaxy density distributions of E+A galaxies at scales of 0.5, 1.5, and 8 Mpc are consistent with that of field galaxies, and inconsistent with that of cluster galaxies. The result implicates that E+A galaxies are not created by a physical mechanism associated with galaxy clusters or the large scale structure.

 \item E+A galaxies have excess number of companion galaxies only at a scale of $<$ 100 kpc compared with the randomly selected field galaxies (Figs \ref{fig:kpc_raw} and \ref{fig:kpc_ratio}). In addition, we found 7 (out of 24) E+A galaxies presented in Fig. \ref{fig:ea7_images} show possible companion/tidally disturbed signatures.
 Based on these results, we propose dynamical interaction with close companion galaxies is the most likely origin of E+A galaxies.

\end{itemize}

\bigskip

\section*{Acknowledgments}

We thank Dr. Ann Hornschemeier for encouraging us to finish this work. We are grateful to Dr. Ani Thakar for his friendly help in downloading the publicly available SDSS data.
 We thank the anonymous referee for many insightful comments, which improved the paper significantly.
    
    Funding for the creation and distribution of the SDSS Archive has been provided by the Alfred P. Sloan Foundation, the Participating Institutions, the National Aeronautics and Space Administration, the National Science Foundation, the U.S. Department of Energy, the Japanese Monbukagakusho, and the Max Planck Society. The SDSS Web site is http://www.sdss.org/.

    The SDSS is managed by the Astrophysical Research Consortium (ARC) for the Participating Institutions. The Participating Institutions are The University of Chicago, Fermilab, the Institute for Advanced Study, the Japan Participation Group, The Johns Hopkins University, Los Alamos National Laboratory, the Max-Planck-Institute for Astronomy (MPIA), the Max-Planck-Institute for Astrophysics (MPA), New Mexico State University, University of Pittsburgh, Princeton University, the United States Naval Observatory, and the University of Washington.



%
%
%

\clearpage

\clearpage

\clearpage

\clearpage

\end{document}